\documentclass[aip,pop,preprint]{revtex4-1}
\renewcommand{\section}[2]{}
\usepackage{graphicx}
\usepackage{amsmath}
\usepackage{amssymb}
\usepackage{bm}
\usepackage{natbib}
\usepackage{color}
\usepackage[normalem]{ulem}

\newcommand{\grad}{\boldsymbol{\nabla}}
\newcommand{\cross}{\boldsymbol{\times}}

\newcommand{\ue}{\ensuremath{\mathbf{u}_{_E}}}

\newcommand{\uek}{\ensuremath{\mathbf{u_E} \boldsymbol{\cdot} \boldsymbol{\kappa}}}

\newcommand{\myemail}{jdahlin@umd.edu}

\begin{document}

\preprint{AIP/123-QED}


\title{
Parallel electric fields are inefficient drivers of energetic electrons in magnetic reconnection
}

\author{J. T. Dahlin}
\email[]{\myemail}
\altaffiliation{Jack Eddy Postdoctoral Fellow}
\altaffiliation{Heliophysics Science Division, NASA Goddard Space Flight Center, Greenbelt, MD 20771, USA}
\author{J. F. Drake}
\altaffiliation{Department of Physics, University of Maryland, College Park, Maryland 20742, USA}
\altaffiliation{Institute for Physical Science and Technology, University of Maryland, College Park, Maryland 20742, USA}

\author{M. Swisdak}

\affiliation{Institute for Research in Electronics and Applied Physics, 
University of Maryland, College Park, Maryland 20742, USA}

\begin{abstract}
We present two-dimensional kinetic simulations, with a broad range of initial 
guide fields, that isolate the role of parallel electric fields ($E_\parallel$) 
in energetic electron production during collisionless magnetic reconnection. 
In the strong guide field regime, $E_\parallel$ drives essentially all of the 
electron energy gain, yet fails to generate an energetic component. We suggest 
that this is due to the weak energy scaling of particle acceleration from 
$E_\parallel$ compared to that of a Fermi-type mechanism responsible for 
energetic electron production in the weak guide-field regime. This result
has important implications for energetic electron production in astrophysical
systems and reconnection-driven dissipation in turbulence.
\end{abstract}


\pacs{52.35.Vd,94.30.cp,52.65.Rr,96.60.Iv}

\maketitle


Magnetic fields are significant reservoirs of energy in many plasmas.
Magnetic reconnection converts that energy into other forms,
principally the thermal and kinetic energy of the surrounding
particles. Of particular interest in many systems is the production of
non-thermal particles with energies much larger than typical of the
ambient medium.  Among the phenomena in which such energetic particle
production occurs are gamma-ray bursts \citep{drenkhahn02a,michel94a},
stellar and solar flares \citep{lin03a}, and magnetospheric storms
\citep{oieroset02a}.  Observations of solar flares, in particular,
demonstrate that the acceleration can be particularly efficient: a
large fraction of the electrons in the flaring region become
non-thermal \citep{krucker10a,oka13a}.  The energy content of this
population is comparable to that of the initial magnetic field.

The mechanisms by which reconnection can foster particle acceleration
are a topic of significant interest
e.g. \cite{hoshino01a,zenitani01a,drake05a,pritchett06a,egedal09a,oka10a,hoshino12a}.
Two specific processes have received the most attention.  The first is
acceleration by electric fields parallel to the local magnetic field
($E_\parallel$) \citep{litvinenko96a,drake05a,egedal12a}. However, the
number of electrons that can be accelerated through this mechanism may
be limited because during magnetic reconnection non-zero
$E_{\parallel}$ typically only occur near X-lines and separatrices.

In the second process \citep{drake06a}, charged particles gain energy
as they reflect from the ends of contracting magnetic islands.  (An
analogous process occurs during the acceleration of cosmic rays by the
first-order Fermi mechanism.)  In contrast to the localization of
$E_{\parallel}$, this can occur wherever there are contracting field
lines, including the merging of magnetic islands in the outflows of
single X-line reconnection
\citep{drake06a,oka10a,drake10a,drake13a,hoshino12a} and in turbulent
reconnecting systems where magnetic field lines are stochastic and
conventional islands do not exist \cite{dahlin15a}.  This mechanism is
therefore volume-filling and can accelerate a large number of
particles.

A recent article \cite{dahlin14a} developed a method for calculating
electron acceleration due to both of these mechanisms as well as
betatron acceleration associated with conservation of the magnetic
moment. This study found that Fermi reflection dominates in reconnection 
where the magnetic fields are roughly antiparallel (see also \citep{guo14a,li15a}),
whereas in guide field reconnection
both Fermi reflection and $E_\parallel$ are important drivers of particle
energization. In another study, Numata et al. \cite{numata15a} found
that $E_\parallel$ drove electron heating in a gyrokinetic system
corresponding to an asymptotically strong guide field, and Wang
\cite{wang16a} found a similar result in a system with a guide field
twice that of the reconnecting component.
In another article \cite{dahlin15a}, we showed that the Fermi
mechanism, which scales like $v_\parallel^2$ compared with
$v_\parallel$ for $E_\parallel$, was the dominant accelerator of
energetic electrons in a system with a guide field equal to the
reconnecting component, even though both mechanisms were equal
contributors to the overall electron energy gain.

In this article, we explore electron heating over the full range of
guide fields, from much smaller to much larger than the reconnecting
field. We find that Fermi reflection is the dominant mechanism in
reconnection with a weak guide field, whereas $E_\parallel$ drives
essentially all of the electron energization in the strong guide field
regime. We present simple models for each mechanism that reveal the
essential physics behind the guide field scaling.  Most significantly
we show that energetic electron production is strongly suppressed in
the strong guide field regime where the parallel electric field
dominates electron energy gain, suggesting more generally that
parallel electric fields are not efficient drivers of energetic
particles in nature.

In order to examine electron energization
we assume a guiding-center approximation relevant for a 
strong guide field \citep{northrop63a,dahlin15a}. In this limit, the evolution of the kinetic energy $\epsilon = (\gamma-1) m_e c^2$ 
of a single electron can be written as:
\begin{equation}
\label{eqn:particle}
\frac{d \epsilon}{d t} = q E_\parallel v_\parallel
  + \frac{\mu}{\gamma}\left( \frac{\partial B}{\partial t} + \ue \boldsymbol{\cdot} \boldsymbol{\nabla} B \right)
  + \gamma m_e v_\parallel^2 (\ue \boldsymbol{\cdot} \boldsymbol{\kappa})
\end{equation}
where $E_\parallel = \mathbf{E} \boldsymbol{\cdot} \mathbf{b}$ is the parallel electric field,
$\mu = m_e \gamma^2 v_\perp^2/2B$ is the magnetic moment, $\ue = c\mathbf{E} \cross \mathbf{B}/B^2$,
and $\boldsymbol{\kappa} = \mathbf{b} \boldsymbol{\cdot} \nabla \mathbf{b}$ is the magnetic curvature.
The velocity components parallel and perpendicular to the magnetic field are
$v_\parallel$ and $v_\perp$, respectively; $\gamma$ is the relativistic Lorentz factor, and
$\mathbf{b}$ is the unit vector in the direction of the local magnetic field.

The first term on the right-hand-side of the equation corresponds to
acceleration by a parallel electric field, which is typically
localized near the reconnection X-line and along separatrices. The
second term corresponds to betatron acceleration associated with $\mu$
conservation in a temporally and spatially varying magnetic
field. Because reconnection releases a system's magnetic energy, this
typically causes electron cooling \cite{dahlin14a}.
The last term corresponds to Fermi reflection of particles from contracting magnetic field lines
\citep{drake06a,drake10a,hoshino12a,dahlin14a}. 
Both $E_\parallel$ and Fermi reflection change the parallel energy 
of the particles, while betatron acceleration changes the perpendicular energy.
The term $\uek$ corresponds to a local field-line contraction: $\uek =
-\dot{\ell}/\ell$ (where $\ell$ is the field-line length) and is linked to 
the conservation of the parallel adiabatic invariant $J_\parallel = \int v_\parallel d\ell$ \cite{drake06a,drake10a}. The guide-field approximation given in Eq.~(\ref{eqn:particle}) is accurate
when electrons are well-magnetized. In the weak-guide field regime, other terms
such as the polarization drift may be significant (compare Li et al.\cite{li15a}). 
However, the polarization drift gives the change in the electron bulk flow 
energy which is typically small for a physical mass ratio.

It is informative to compare Eq.~\ref{eqn:particle} to the evolution of the magnetic
energy:
\begin{equation}
\label{eqn:mag1}
\frac{\partial}{\partial t} \frac{B^2}{8\pi} 
 +\grad \cdot \mathbf{S}=
- \mathbf{E} \cdot \mathbf{J} 
\end{equation}
where $\mathbf{S} = c\mathbf{E} \times \mathbf{B}/4\pi$ is the Poynting flux and the
displacement current has been neglected in Amp\`ere's law.
The current $\mathbf{J}$ may be separated into parallel and perpendicular components
so that Eq.~(\ref{eqn:mag1}) becomes:
\begin{equation}
 \label{eqn:mag2}
\frac{\partial}{\partial t} \frac{B^2}{8 \pi} + \boldsymbol{\nabla} \boldsymbol{\cdot} \left(\frac{B^2}{4\pi} \ue \right)
= - E_\parallel J_\parallel 
-  \mathbf{E} \cdot \left[\frac{c\mathbf{B}}{B^2} \cross \left(\frac{\mathbf{B} \cdot \grad \mathbf{B}}{4\pi} 
-\frac{\grad B^2}{8\pi}\right) \right]
\end{equation}
This equation may be rearranged into the following useful form:
\begin{equation}
  \label{eqn:mag3}
\frac{\partial}{\partial t} \frac{B^2}{8\pi}
+ \grad\boldsymbol{\cdot}\left(\frac{B^2}{8\pi}\ue \right) =
- E_\parallel J_\parallel
-\frac{B^2}{8\pi}\grad\boldsymbol{\cdot}\ue
- (\ue \boldsymbol{\cdot}\boldsymbol{\kappa}) \frac{B^2}{4\pi}
\end{equation}
The second term on the left-hand side is the divergence of magnetic
energy flux, which vanishes in a volume-integration.
The first term on the right-hand side has a clear analogue in the 
$E_\parallel$ term in Eq.~(\ref{eqn:particle}). 
The third term corresponds to the mechanical work done by the magnetic tension force
($\boldsymbol{\kappa} B^2/4\pi$).  
This term contains the field-line contraction $\uek$ and is therefore related to the
Fermi reflection term in Eq.~(\ref{eqn:particle}). 
Beresnyak and Li\cite{beresnyak16a} have also noted the link between energy conversion 
via the tension force and first-order Fermi acceleration via the curvature drift.
The $\grad\boldsymbol{\cdot} {\bf u}_E$ term describes the
change in magnetic energy associated with compression or expansion. 


We explore particle acceleration in reconnection via simulations using
the particle-in-cell (PIC) code {\tt p3d}
\citep{zeiler02a}. Particle trajectories are calculated using the
relativistic Newton-Lorentz equation, and the electromagnetic fields
are advanced using Maxwell's equations. The time and space coordinates
are normalized, respectively, to the proton cyclotron frequency based
on the reconnecting magnetic field $\Omega_{ci}^{-1} = m_i c/eB_{x0}$
and the proton inertial length $d_i = c/\omega_{pi}$.  The typical grid cell
width is $\Delta = d_e/8$ where $d_e = d_i \sqrt{m_e/m_i}$ is the
electron inertial length, and the time step is $dt =
\Omega_{ci}^{-1}/200 = \Omega_{ce}^{-1}/8$, where $\Omega_{ce} =
(m_i/m_e)\Omega_{ci}$ is the electron cyclotron frequency.  The domain
size is $51.2 d_i \times 25.6 d_i$, and we vary the guide field $b_g
\equiv B_{z0}/B_{x0} \in [0,0.2,0.5,1.0,2.0,4.0]$. For $b_g = 4.0$,
$\Delta = d_e/16$ and $dt = \Omega_{ci}/400$.  In terms of electron
Larmor radius, $\Delta/\rho_e \in
[0.177,0.180,0.198,0.25,0.395,0.364]$.

We use an artificial proton-to-electron mass ratio $m_i/m_e = 25$ to
reduce the computational expense.  All simulations use at least $200$
particles per cell. The initial electron and proton temperatures are
isotropic, with $T_e = T_i = 0.25m_i c_A^2$, and the initial density
$n_0$ and pressure $p$ are constant so that $\beta_x = 8\pi p/B_{x0}^2
= 0.5$. The speed of light is $c = 3 c_A \sqrt{m_i/m_e}$, where
$c_{A}=B_0/\sqrt{4\pi m_i n_0}$ is the Alfv\'en speed based on the
reconnecting component of the magnetic field.

All simulations are initialized with a force-free configuration and
use periodic boundary conditions. This is chosen as the most generic
model for large-scale systems such as the solar corona where the
density jump between the current layer and upstream plasma is not
expected to be important.  The magnetic field is given by $B_x =
B_{x0} \tanh(y/w_0)$ and $B_z = \sqrt{(1+b_g^2)B_{x0}^2-B_x^2}$,
corresponding to an asymptotic guide field $B_{z0}= b_g B_{x0}$.  We
include two current sheets at $y=L_y/4$ and $3L_y/4$ to produce a
periodic system, and $w_0 = 1.25d_e$.  This initial configuration is
not a kinetic equilibrium, which would require a temperature
anisotropy \cite{bobrova01a}, but is in pressure balance.

\begin{figure}
 \includegraphics{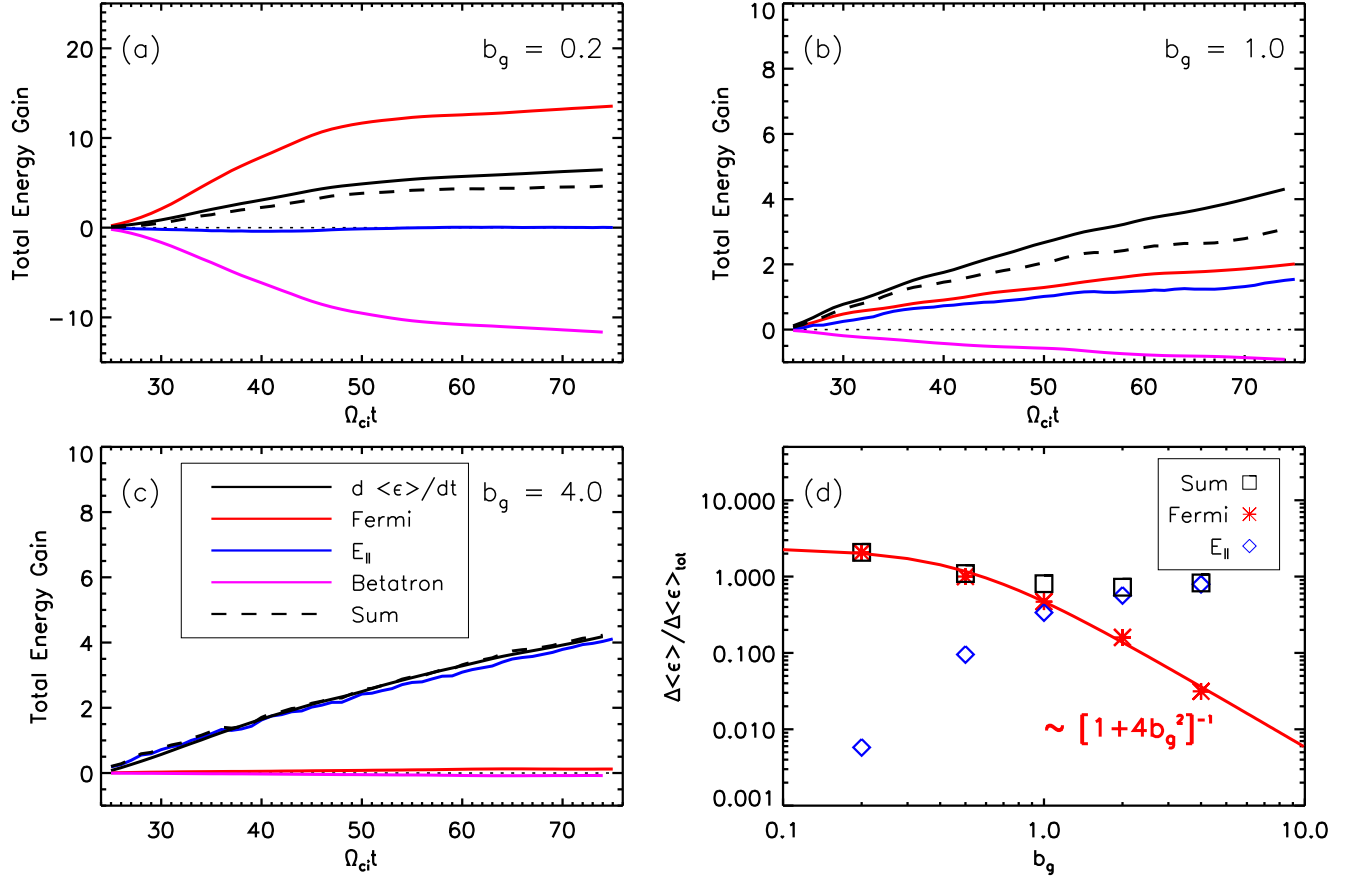}
 \caption
{(a)-(c) Cumulative electron heating due to Fermi reflection (red), $E_\parallel$ (blue) and
betatron acceleration (magenta) for three different values of $b_g$. (d) Total electron energy gain due to Fermi and $E_\parallel$ as
a function of guide field.
\label{fig:heat_all}}
 \end{figure}

Reconnection begins from noise via the tearing instability, generating
magnetic islands which grow and merge. Reconnection evolves
nonlinearly until we halt the simulations before the two current
sheets significantly interact.  Panels (a-c) of 
Figure \ref{fig:heat_all} show the
cumulative electron energy gain due to the three mechanisms in
Eq.~(\ref{eqn:particle}). Bulk electron heating is calculated via the
methods discussed in an earlier work \cite{dahlin14a}. In all cases
the terms given in Eq.~(\ref{eqn:particle}), indicated by the dashed 
black line, adequately capture the total energy gain of the electrons 
(solid black line). The difference is due to several small terms, such
as the polarization drift, that were omitted from this equation.

\begin{figure}
 \includegraphics[width=6in]{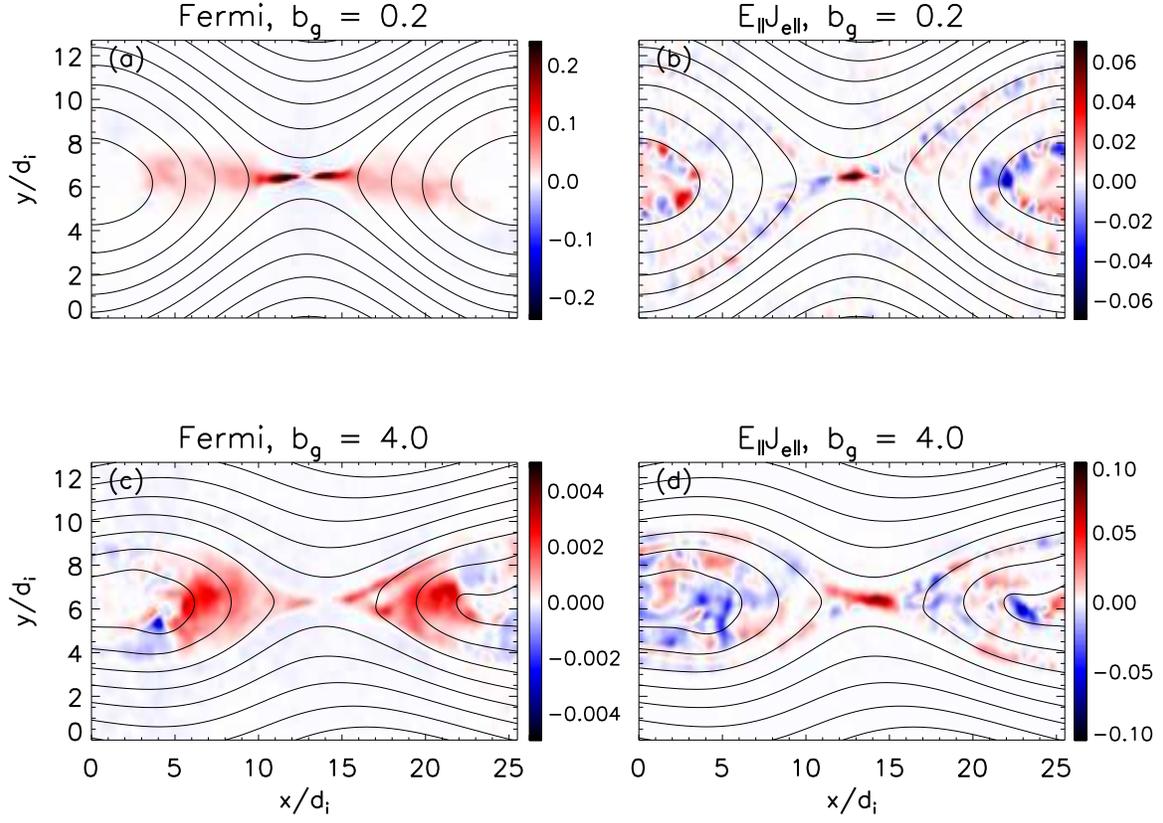}
 \caption
{(a)(c) Fermi acceleration for two values of the guide field. (b)(d) $E_\parallel J_{e\parallel}$
acceleration for two values of the guide field, where $J_{e\parallel}$ is the parallel component of
the electron current.
\label{fig:heating}}
 \end{figure}

In the weak guide field case (Fig.~\ref{fig:heat_all}a), the largest terms are Fermi
reflection (positive) and betatron acceleration (negative) which
partially cancel to produce most of the electron energy gain. The
contribution of parallel electric fields is negligible. The betatron
term is larger than in the previous study (Dahlin et
al.\cite{dahlin14a}) because it scales proportionally with the plasma
$\beta$, which is significantly larger here ($\sim 1$ vs. $\sim 0.2$). 
In the simulation with $b_g = 4$ (Fig.~\ref{fig:heat_all}c), all terms but
$E_\parallel$ have negligible contributions to electron energy gain, a
result similar to Wang et al., 2016 \cite{wang16a}.  In the
intermediate case $b_g = 1$, both Fermi reflection and $E_\parallel$
are important, as was reported in \cite{dahlin14a}.  Note that the
total electron energy gain is about $\sim 50\%$ larger in the $b_g =
0.2$ case than in the $b_g = 4.0$ case.  As $b_g$ increases the system
becomes less compressible, the consequences of which are discussed
further below.  Figure \ref{fig:heating} shows the late-time spatial
distribution of electron heating in simulations with $b_g = 0.2, 4.0$.
Although the location of the
heating is similar for both cases, the magnitude changes with
Fermi reflection decreasing by around a factor of $50$.
The energization due to Fermi and $E_\parallel$ (normalized to the
total electron energy gain) is shown versus the guide field in
Fig.~\ref{fig:heat_all}d. The Fermi term is greater than unity
for small $b_g$ due to the large cancellation with the betatron term.
It is clear that the contribution of Fermi reflection falls off
rapidly with increasing guide field, and vice-versa for $E_\parallel$.

\begin{figure}
 \includegraphics{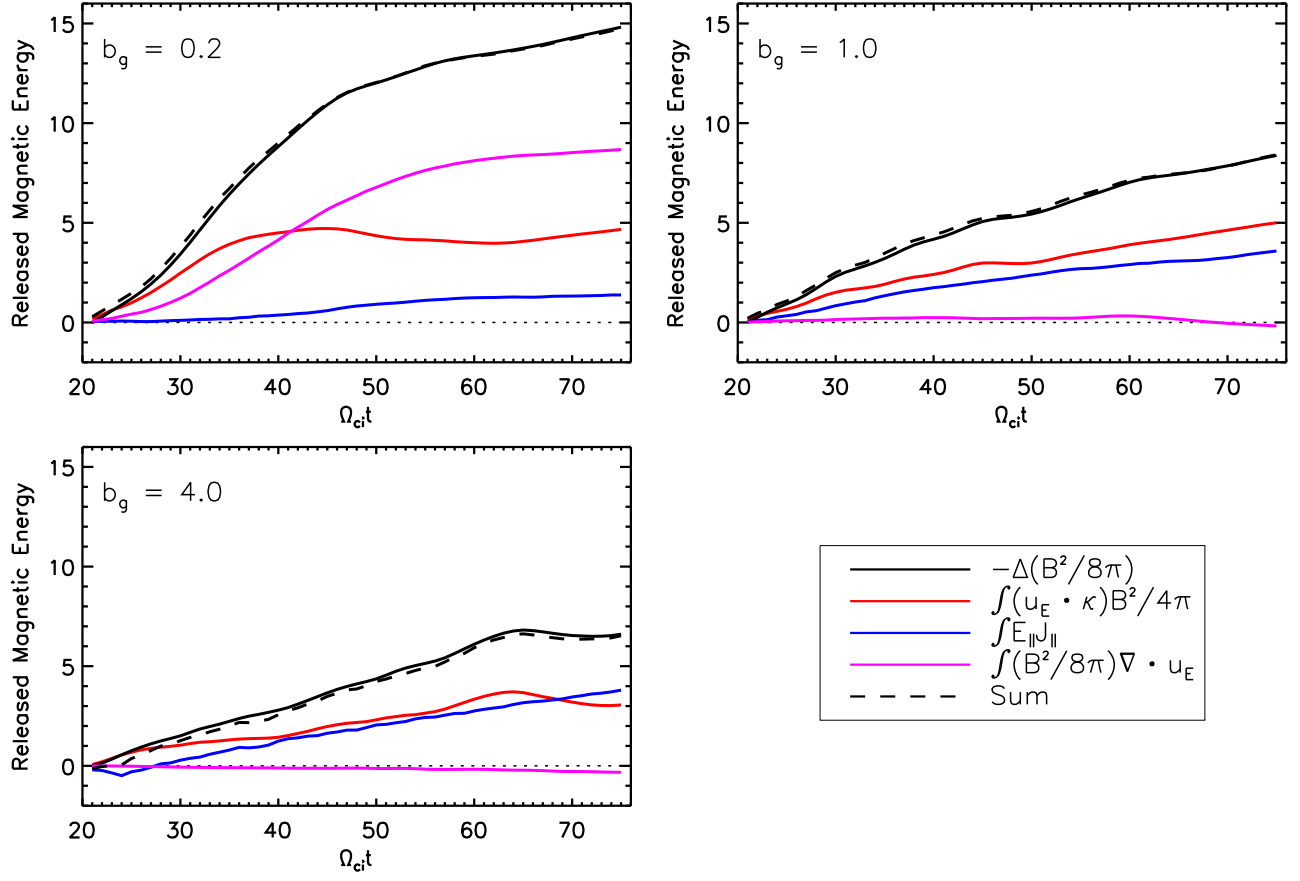}
 \caption{Time dependence of the spatially integrated mechanisms for magnetic energy release given in Eq.~\ref{eqn:mag2} for guide fields $b_g=0.2$, $1.0$ and $4.0$.}
\label{fig:release_all}
 \end{figure}

Given the dramatic change in the mechanisms driving electron
acceleration with increasing guide field, it is informative to also
explore how magnetic energy is being released during the same
transition. In Fig.~\ref{fig:release_all} we show the time dependence
of the spatially integrated rates of magnetic energy release for the
three terms on the RHS of Eq.~\ref{eqn:mag2}. For weak guide field
($b_g = 0.2$) the dominant terms are from field-line expansion and 
magnetic tension with the contribution from $E_\parallel$ being
small. For larger guide fields ($1.0$ and $4.0$) the curvature and
$E_\parallel$ terms are comparable and the compression term is
negligible -- the guide field clearly suppresses compression. Since
electron energization from Fermi relection is so weak at high guide
field, the continued importance of magnetic tension in releasing magnetic
energy might be surprising. However, it is well known that the outflow
exhaust velocity remains at $c_A$ even when $b_g$ is large
\citep{lin92a} so the ion bulk flow carries much of the magnetic
energy released.

Fermi acceleration is driven by the reflection of a charged particle from a field line expelled by the exhaust with a velocity 
$c_A$, where $c_A$ is the Alfv\'en velocity based on the reconnecting component.
The energy gain due to a single reflection is given by:
\begin{equation*}
\Delta \epsilon \approx 2 m c_A \mathbf{x} \cdot v_\parallel \mathbf{b} = 2 m v_\parallel c_A \frac{B_x}{B}
\end{equation*}
The time between reflections is:
\begin{equation*}
\Delta t \sim L/v_{x} \sim LB/(v_\parallel B_x) 
\end{equation*}
where $L$ is the characteristic island length. This yields an energization rate \cite{drake06a}:
\begin{equation} 
\label{eqn:uekscaling}
\dot{\epsilon} \sim m v_\parallel^2 \frac{2 c_A}{L} \frac{B_x^2}{B^2} \propto \frac{1}{1+B_z^2/B_x^2}
\end{equation}
Thus, in the strong guide field regime $B_z \gg B_x$ the scaling is $\dot{\epsilon}
\propto b_g^{-2}$. Only the component of the electron parallel
velocity along the direction of contracting magnetic field contributes
to energy gain so there is less energy gain per
reflection. Additionally, the time between reflections is greater in
the large guide-field limit because the reflection frequency is
proportional to the in-plane velocity $v_x \sim v_\parallel/b_g$. The
diminished efficiency of Fermi reflection can also be seen in the
scaling of the magnetic curvature $\boldsymbol{\kappa} \sim
(\mathbf{B} \cdot \grad \mathbf{B})/B^2 \propto B_xB_y/\delta B^2 \propto
B_x^2/\delta B^2$, where $\delta$ is the characteristic width of the exhaust and $B_y/B_x \sim 0.1$ is linked to the aspect ratio of the diffusion region. In the strong guide field regime, the reconnected field lines
are elongated in the out-of-plane direction so that advection in the
$x-y$ plane does little to change their overall length.
In the weak guide field regime, $B \approx B_x$ and Eq.~\ref{eqn:uekscaling} is independent of $B_z/B_x$. 
The curve $\sim[1+4b_g^2]^{-1}$ 
(solid red line in Fig.~\ref{fig:heat_all}) describes the scaling of the heating from Fermi reflection very well (the factor of 4 yields the best fit).

In an earlier study\cite{dahlin14a}, we showed that the electron energy 
gain due to $E_\parallel J_\parallel$ occurs around the diffusion region 
near the X-line (the positive and negative values in the blotchy patches 
distributed along separatrices and in the islands in Fig.~\ref{fig:heating}
do not drive net heating). Within the diffusion region where the plasma is not 
frozen-in, the reconnection electric field and out-of plane current drive 
the energy conversion: \begin{equation} E_\parallel J_\parallel \sim 
\frac{J_z B_z}{B} \frac{E_z B_z}{B} \end{equation} 
In the weak guide-field limit this
term is only significant in the region where $B_z/\sqrt{B_x^2+B_y^2} >
1$.  The area where this inequality is satisfied is given by
$b_g^2\delta_x\delta_y$, where $\delta_x$ and $\delta_y$ are the scale
lengths of $B_y$ and $B_x$ in the $x$ and $y$ directions,
respectively. Thus, the fraction of the diffusion region where
$J_\parallel E_\parallel$ is significant scales as $b_g^2$ in the weak guide field
regime $b_g \ll 1$. This simplified description neglects important 
antiparallel dynamics (such as the Hall fields and meandering particle orbits) and
does not completely match the simulation results, which are more consistent with
$b_g^3$, but serves to
illustrate the scaling of $E_\parallel$ heating 
with increasing guide field.
When the guide field is large ($b_g \gtrsim 1$), the diffusion region is defined as the
region where $E_\parallel$ is non-zero and drives $J_\parallel$. In this limit, the
entire diffusion region contributes to dissipation through $J_\parallel E_\parallel$.

\begin{figure}
 \includegraphics{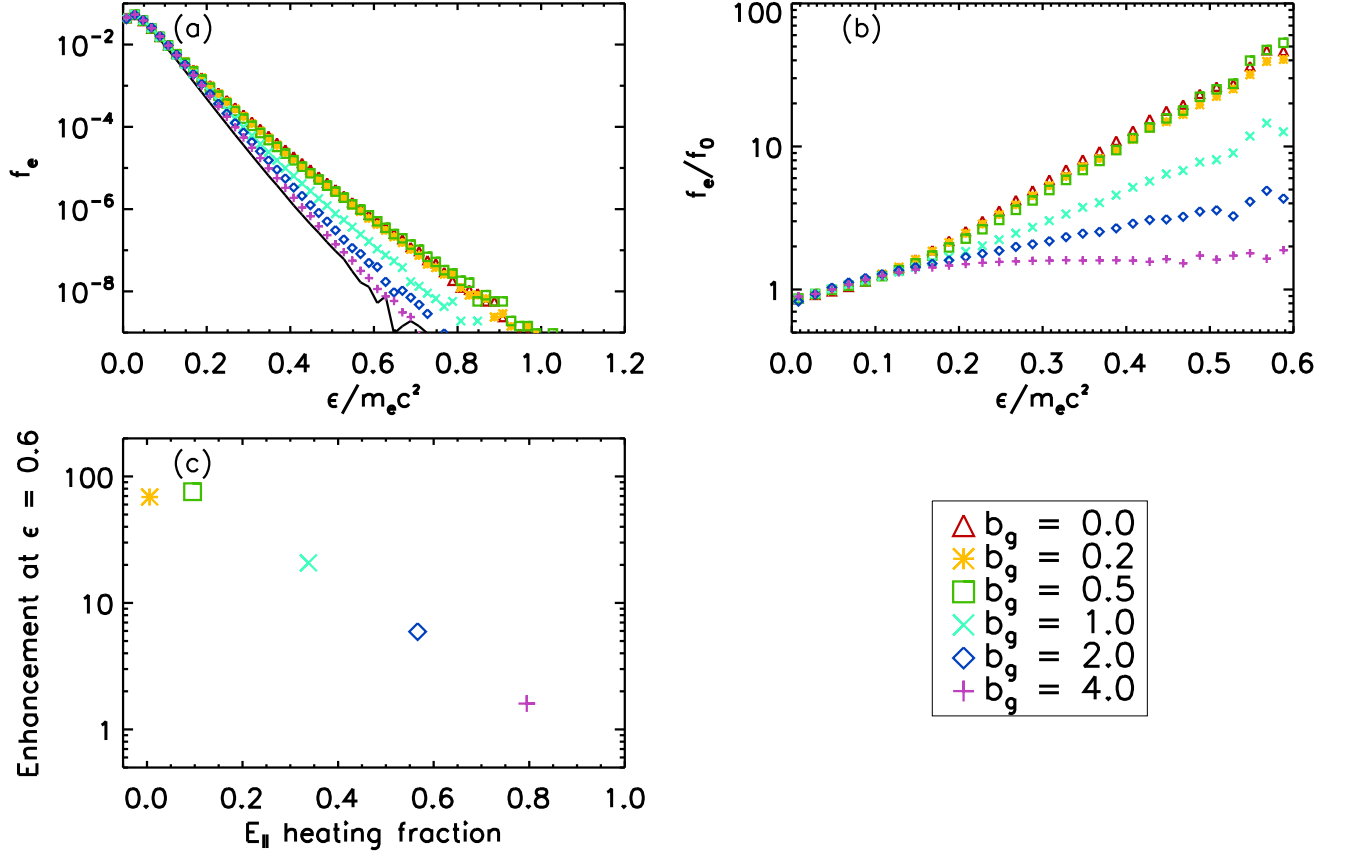}
 \caption
{(a) Electron energy spectra at $\Omega_{ci}t = 75$.
(b) Enhancement relative to initial spectrum $f_e(\epsilon,t=75)/f_e(\epsilon,t=0)$.
(c) Energetic electron enhancement ($\epsilon = 0.6$) versus fraction of electron heating 
due to $E_\parallel$. The $E_\parallel$ heating for $b_g = 0$ is not calculated due to the poor applicability of the 
guiding-center model.
\label{fig:spectra2d_guide}}
 \end{figure}

Energy spectra at late time $\Omega_{ci}t = 75$ 
(Fig.~\ref{fig:spectra2d_guide}a) 
reveal that the production of energetic electrons in reconnection with 
strong guide fields is nearly completely eliminated. The normalized spectra
(Fig.~\ref{fig:spectra2d_guide}b) showcase the enhancement 
in energetic electrons, given by 
$f_e(\epsilon,t=75)/f_e(\epsilon,t=0)$.
There is a substantial enhancement (a factor of $\sim 50$) at $\epsilon =
0.6 m_ec^2$ for the systems with $b_g < 1$, whereas for the system
with $b_g = 4$ the enhancement is $< 2$, and is approximately
independent of $\epsilon$ for $\epsilon > 0.2$. 
Note that in these simulations, the contribution 
of the energetic population to the total energy is small, so that the 
overall conversion of magnetic to electron energy is only weakly dependent 
on the guide field (see Fig.~\ref{fig:heat_all}a-c).

Panel (c) shows
the energetic electron enhancement at $\epsilon = 0.6$ versus the
electron heating fraction due to $E_\parallel$ (the same quantity is shown 
in Fig.~\ref{fig:heat_all}d). It is clear that energetic electron
production diminishes rapidly as the $E_\parallel$ contribution
increases. 
According to Eq.~(\ref{eqn:particle}), energy gain from $E_\parallel$
is a weaker function of energy ($\sim v_\parallel$) than that due
to the Fermi mechanism ($\sim v_\parallel^2$). Whereas the Fermi mechanism
preferentially accelerates energetic particles to generate a nonthermal tail, 
parallel electric fields distribute energy more evenly and drive bulk
heating. This result suggests that parallel electric fields are, in general, 
inefficient drivers of energetic electrons.



These results have broad implications for electron energization and
heating in reconnecting systems where the reconnecting component of
the magnetic field is small compared to the guide field. For example,
this suggests that $E_\parallel$ is the most important
reconnection-driven heating mechanism in turbulent systems with
$\delta B/B \ll 1$, consistent with the results of gyrokinetic
simulations \cite{numata15a}. Notably, such reconnection is not likely
to produce large fluxes of energetic particles. In
reconnecting systems with broad current layers where the strength of
the effective reconnecting magnetic field increases with time as
stronger magnetic fields convect toward the reconnection site (as may
be the case in a solar flaring current sheet) \citep{drake14a}, the
rate of production of the most energetic electrons should increase
rapidly with time.

A notable limitation of this study is the artificial `2D' constraint
(equivalent to $\partial/\partial z = 0$).  A recent study\cite{dahlin15a}
 showed that electron energization is greatly enhanced in a three-dimensional
system with a strong guide field where reconnection becomes turbulent
and high-energy electrons are able to move freely to sample regions
where energy release is taking place.  However, in three-dimensional
systems with a weak guide field, transport is diminished and there is
little enhancement (Dahlin et al., in preparation). This suggests that
the most efficient energetic electron production might occur for
$b_g \sim 1$ where the Fermi mechanism and three-dimensional dynamics
are both important. A complete theory of the production of energetic
electrons from reconnection must incorporate the role of the guide
field in transport and in the strength of the various energy drive
mechanisms.

\clearpage
\clearpage

\clearpage

\clearpage

This work has been supported by NSF Grant
PHY1500460 and NASA grants
NNX14AC78G, NNX14AF42G, and DOE grant DEFG0293ER54197.
J.T.D. acknowledges
support from the NASA LWS Jack Eddy Fellowship
administered by the University Corporation for Atmospheric
Research in Boulder, Colorado.
Simulations were carried out at
the National Energy Research Scientific Computing Center.

\bibliography{../paper}

\end{document}